\documentclass[article,showpacs,preprintnumbers,amsmath,amssymb]{revtex4}
\usepackage{graphicx}% Include figure files
\usepackage[usenames]{color}
\usepackage{subfigure}

\begin{document}

\title{Leptoquarks signals in KM$^3$ neutrino telescopes}

\author{Ismael.Romero}
 \affiliation{Departamento de F\'{\i}sica,
Universidad Nacional de Mar del Plata \\
Funes 3350, (7600) Mar del Plata, Argentina}

\author{Oscar A.Sampayo}
\email{sampayo@mdp.edu.ar}

 \affiliation{Departamento de F\'{\i}sica,
Universidad Nacional de Mar del Plata \\
Funes 3350, (7600) Mar del Plata, Argentina}

\begin{abstract}
Leptoquarks are predicted in several extensions of the Standard
Model (SM) of particle physics attempting the unification of the
quark and lepton sectors. Such particles could be produced in the
interaction of high energy neutrinos with matter of the Earth. We
investigate the effects of this particles on the neutrino flux to be
detected in a kilometer cubic neutrino telescope such as IceCube. We
calculate the contribution of leptoquarks to the neutrino-nucleon
interaction and, then, to the angular observable $\alpha(E)$
recently proposed in order to evaluate detectable effects in
IceCube. Our results are presented as an exclusion plot in the
relevant parameters of the leptoquarks physics.
\end{abstract}


\maketitle

\section{Introduction}
%%%%%%%%%%%

%%%%%%%%%%%%%%

The Standard Model (SM) for the  elementary particles interactions
has been successfully tested at the level of quantum corrections. In
particular high precision and collider experiments have tested the
model and have placed the border line for new physics effects at
energies of the order of $1 \rm{TeV}$ \cite{pdg}. On the other hand,
new physics effect in the neutrino sector have recently received an
important amount of experimental information coming from flavor
oscillation \cite{oscneu}. This fact is the first evidence of
neutrino masses different from zero, and hence, of physics beyond
the SM. In this way, the neutrino sector and in particular
neutrino-nucleon interactions, could be the place where new physics
may become manifest again.

Although the SM has been successful to describe the world at short
distances, as a low energy effective theory of phenomena at higher
scales, it leaves several open questions, e.g.: it does not predict
the number of families and the fermions masses,  has  several free
parameters, the mass generation mechanism through the Higgs boson,
where its mass is not predicted, is untested  and still leaves open
the hierarchy problem. In these conditions, it is believed that we
should have some kind of physics beyond the SM, which is called New
Physics (NP)\cite{pdg}. The search of NP proceeds mainly through the
comparison of data with the SM predictions. The experimental way to
look for NP effects in a model independent fashion is to construct
observables that can be affected by this new physics and then
compare the  measurements with the mentioned SM expectation. Certain
types of NP can already be present at the TeV scale and could
participate in neutrino-nucleon interactions. Hence, these NP
effects could possibly become apparent in neutrino telescopes. This
detectors are able to explore the high energy neutrino-nucleon
collision, reaching centre-of-mass energies orders of magnitude
above those of man made accelerators. Although having large
uncertainties on the beam composition and fluxes, cosmic ray
experiments present a unique opportunity to look for new physics at
scales far beyond the TeV when energetic cosmic and atmospheric
neutrinos interact with the nucleons of the Earth. In this sense an
observable recently defined, which is weakly dependent of the
initial flux \cite{alfa}, was used to bound physics beyond the
standard model \cite{alfanp}. In this work we are interested in to
study the effects originated in leptoquark physics on this
observable and other related. In particular we use the angular
observable $\alpha(E)$ (and the related observable $\eta(E)$) to
bound leptoquarks effects. Our results are comparable with the one
obtained by using the inelasticity as observable \cite{leptogarcia}.
In our calculation we use the neutrinos flux arriving to the
detector after through the Earth, for the all energy range. Thus,
the earth stop the atmospheric muons, vanishing the corresponding
background. Up-going muon events from CC $\nu_{\mu}$ interactions
produce an energetic muon traversing the detector. The selection of
these events eliminate the background of atmospheric muons. This is
the traditional observation mode. Simulations, baked by AMANDA data,
indicate that the direction of muons can be determined with sub
degree accuracy and their energy measured to better than $30\%$ in
the logarithm of the energy. The important advantage of this mode is
the angular sub-degree resolution which is a fundamental fact for
the definition of the observable $\alpha(E)$. In the other hand, as
it was recognized by the authors of
Ref.\cite{leptogarcia,bh-anchor}, if we take as detection volume for
contained events the instrumented volume (for IceCube roughly
1km$^3$) IceCube we will have sufficient energy resolution to
separately assign the energy fractions in the muon track and the
hadronic shower allowing the determination of the inelasticity
distribution and the neutrino energy. Recently the possibility to
measure the inelasticity distribution  was used to study the
possibility to put bounds to new effects coming from leptoquarks or
Black-Hole production over kinematics regions ever tested
\cite{leptogarcia,bh-anchor}. In our particular case the possibility
of measured independently the muon energy and the hadronic shower
energy will allow us a reasonable $\nu_{\mu}$-energy determination.
In the following we take the uncertainties in the $\nu_{\mu}$-energy
as $\Delta log_{10}E=0.5$.

How we will explain below in this work we use the guaranteed
atmospheric muon neutrino flux added to a isotropic cosmic neutrino
flux, lower than the AMANDA bound but higher than the Waxman-Bachall
level.

The existence of families of quarks and leptons suggest a possible
link between these two sectors \cite{lepto1}. Many theories, like
composite models, technicolor, and grand unified theories, predict
the existence of new particles, called leptoquarks, that mediate
quark-lepton transitions \cite{lepto2}. It is important to realize
that simultaneous trilinear coupling of the leptoquark to a purely
hadronic channel is excluded in order to avoid too fast barion decay
\cite{lepto3}. In this work, in order to illustrate the behavior of
our observable, $\alpha(E)$, with this kind of new physics we have
considered the simple case of $SU(2)$-singlet scalar leptoquark
$\mathcal{S}$ coupled to the second family, which interact with
quarks and leptons through the lagrangian
\begin{eqnarray}\label{laglq}
{\cal L}_{LQ}=(g_L \bar Q^c_L i \tau_2 L_L + g_R \bar c^c_R
\mu_R)\mathcal{S}
\end{eqnarray}

\begin{figure}[t!]
\centering
\includegraphics[width=4in]{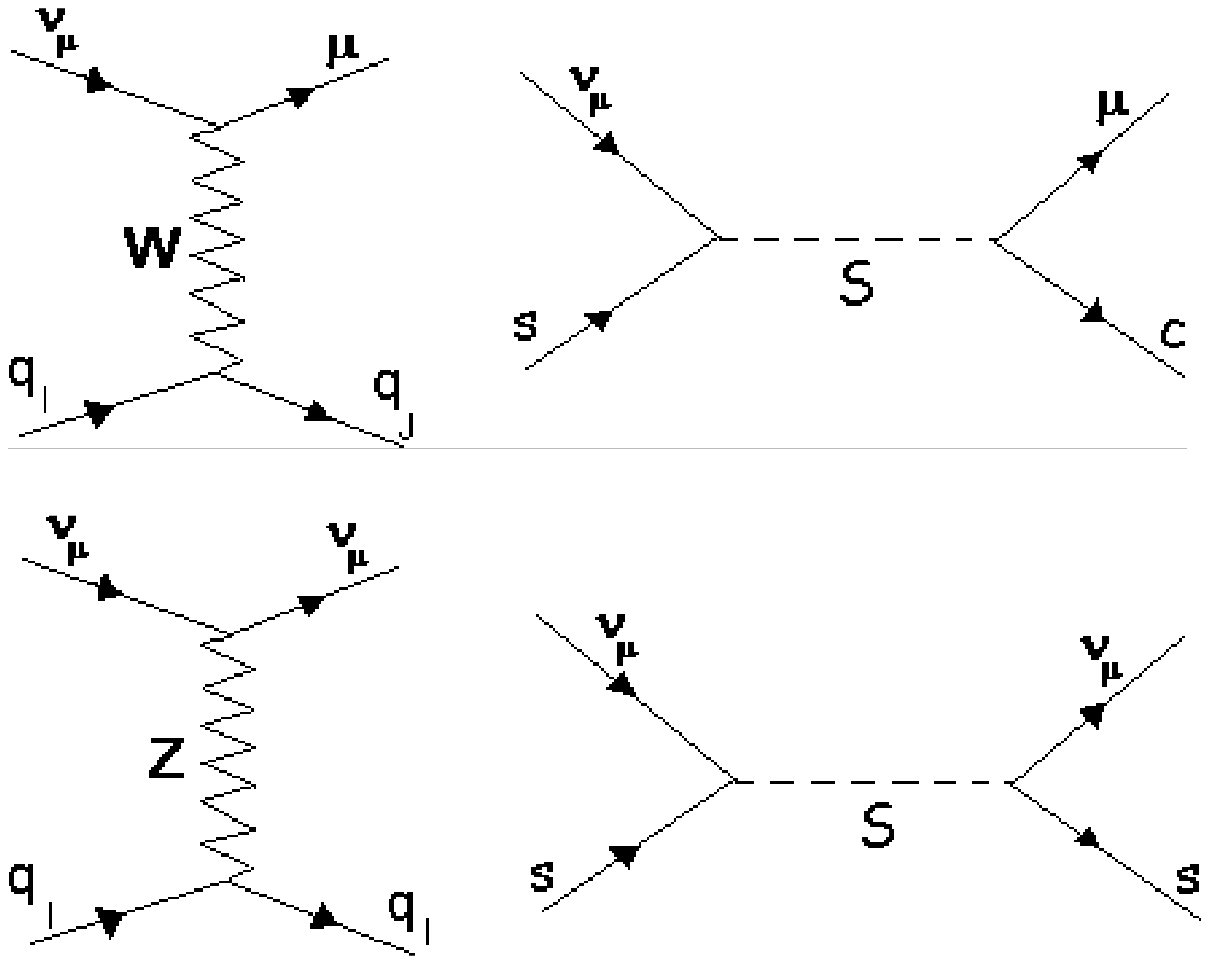}
\caption{\label{fig:diag}  Diagrams contributing to the
 neutrino-nucleon
cross section.}
\end{figure}

where $Q=(c,s)^t$ and $L=(\nu_{\mu},\mu)^t$ are the quarks and
leptons $SU(2)$ left-handed doublets, $c_R$ and $\mu_R$ are the
right singlets, and $g_L$ and $g_R$ the corresponding coupling
constant.

We are interested in the second family since the directions of the
produced muons can be determined with high accuracy.

In figure~\ref{fig:diag} we show beside the SM contribution for
charged and neutral current the leptoquark relevant diagram, where
the corresponding cross section for charged and neutral current are
given by:

\begin{eqnarray}
\frac{d\sigma^{CC}_{LQ}}{dxdy}&=&\frac{g_L^2}{32 \pi}(g_L^2+g_R^2)
\frac{\hat{s}}{(\hat{s}-M^2_{LQ})^2+(\Gamma M_{LQ})^2}  s(x,M_{LQ})
\nonumber
\\
\frac{d\sigma^{NC}_{LQ}}{dxdy}&=&\frac{g_L^4}{32 \pi}
\frac{\hat{s}}{(\hat{s}-M^2_{LQ})^2+(\Gamma M_{LQ})^2}  s(x,M_{LQ})
\end{eqnarray}
where $\hat{s}=x S$, $S=2 M_{proton} E_{\nu}$ and the Leptoquark
width is $\Gamma=(M_{LQ}/16 \pi) (2 g_L^2+g_R^2)$. On the other hand
the corresponding SM cross section reads for charged current

\begin{equation}\label{disfsigcc}
\begin{split}
 \frac{d\sigma^{CC}}{dxdy}=\frac{G_F^2 s}{\pi}\left(
\frac{M_W^2}{(Q^2+M_W^2)} \right)^2x[Q^{CC}+(1-y)^2 \bar Q^{CC}],
\end{split}
\end{equation}

and for the neutral current

\begin{equation}\label{disfsignc}
\begin{split}
\frac{d\sigma^{NC}}{dxdy}=\frac{G_F^2 s}{\pi}\left(
\frac{M_Z^2}{Q^2+M_Z^2} \right)^2
\sum_{i=U,D}x[g^{i2}_L (Q^{i}+(1-y)^2 \bar Q^{i})  \\
+g^{i2}_R( \bar Q^{i}+(1-y)^2 Q^{i})],
\end{split}
\end{equation}

where the quark combinations $\bar Q^{CC}$, $Q^{CC}$, $\bar Q^i$ and
$Q^i$ for a isoscalar target are given in \cite{alfa,gandhi} and
$g^U_L=1/2-2x_W/3$, $g^D_L=-1/2+ x_W/3$, $g^U_R=-2 x_W/3$,
$g^D_R=x_W/3$, $c_W=\cos \theta_W$, $x_W=\sin^2 \theta_W$.

\begin{figure}
\centering
\includegraphics[angle=270,width=3in,bb= 180 180 550 680]{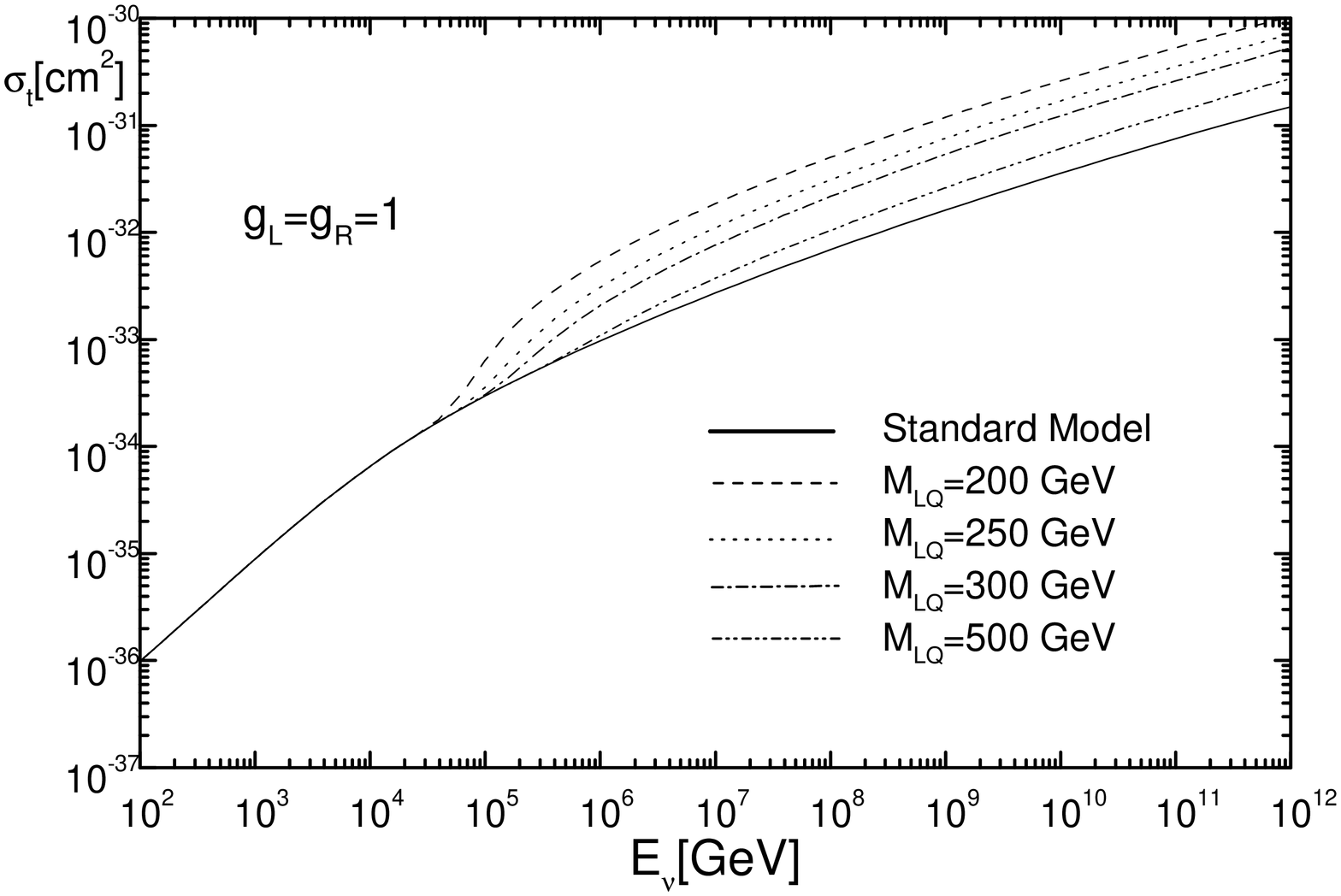}
\caption{\label{fig:sigtot}  Total cross section for the SM
 and for different values of the leptoquark mass $M_{LQ}$
and for $g_L=g_R=1$.}
\end{figure}

In Fig. \ref{fig:sigtot} we show the behavior of the total cross
section ($\sigma^t(E)=\sigma^{CC}(E)+\sigma^{NC}(E)$) with the
neutrino energy for different values of $M_{LQ}$ and the couplings
$g_L=g_R=1$. We can appreciate a disagreement with the SM
predictions, due to the leptoquark contribution for values of
$E_{\nu}$ where the leptoquark can be on shell.

\section{The surviving neutrino flux}

The surviving flux of neutrinos of energy $E$, with inclination
$\theta$ with respect to nadir direction, $\Phi(E,\theta)$, is the
solution of the complete transport equation \cite{nicolaidis}:
\begin{equation}\label{ecuaciontransporte1}
\frac{d \ln\Phi(E,\tau')}{d\tau'}=-\sigma^t(E)+\int_E^{\infty} dE'
\frac{\Phi(E',\tau')}{\Phi(E,\tau')} \; \frac{d\sigma^{NC}}{dE},
\end{equation}
where the first term correspond to absorption effects and the second
one to the regeneration. Here,  $0<\tau^{'}<\tau(\theta)$ where
$\tau=\tau(\theta)$ is the number of nucleons per unit area in the
neutrino path through the Earth,

\begin{equation}\label{tau}
\tau(\theta)=N_A \int_0^{2 R_{\rm E}\cos\theta} \rho(z) dz,
\end{equation}

$N_A$ is the Avogradro number, $R_{\rm E}$ is the radius of the
Earth, $\theta$ is the nadir angle taken from the down-going normal
to the neutrino telescope and the earth density $\rho(r)$ is given
by the preliminary reference earth model \cite{premm}. In order to
find a solution for this equation we make the following
approximation \cite{ralston}: we replace the fluxes ratio inside the
integral of the second member by the ratio of fluxes that solve the
homogeneous equation (i.e., only considering  absorption effects)
\begin{equation}\label{ecuaciontransporte2}
 \frac{\Phi(E',\tau')}{\Phi(E,\tau')} \;\; \rightarrow \;\;
 \frac{\Phi_0(E',\theta)}{\Phi_0(E,\theta)} e^{-\Delta(E',E) \tau'}
\end{equation}
where $ \Delta(E',E)=[\;\sigma^t(E')-\sigma^t(E)\;] $ and
$\Phi_0(E',\theta)$ is the initial flux at the earth surface. How we
explain later, we will use the initial flow given by the sum of the
atmospheric flux, with a well-known angular dependency, with a
diffuse and isotropic cosmic flux.
 Thus,
integrating the transport equation we have
\begin{equation}\label{transporte}
\Phi(E,\theta)=\Phi_0(E,\theta) e^{-\sigma_{eff}(E,\theta)
\tau(\theta)},
\end{equation}
where
\begin{equation}
\sigma_{eff}(E,\theta)=\sigma^t(E) - \sigma^{reg}(E,\theta),
\end{equation}
with
\begin{equation}
\sigma^{reg}(E,\theta)=\int_E^{\infty} dE'  \;
\frac{d\sigma^{NC}}{dE} \left(
\frac{\Phi_0(E',\theta)}{\Phi_0(E,\theta)} \right) \left(
\frac{1-e^{-\Delta(E',E) \tau(\theta)} }{ \tau(\theta) \Delta(E',E)}
\right).
\end{equation}

It is important to mention that the solution of the transport
equation, Eq.~(\ref{transporte}) is the first, but quite accurate,
approximation of the iterative method showed in Ref.\cite{perrone}.

%%%%%%%%%%%%%%%%%%%%%%%%%%%%%%%%%%%%%%%%%%%%%%%%%%%%%%%%%%%%%%%

\section{The Observable {\bf $\alpha(E)$}}
A kilometer cubic neutrino telescope as IceCube is capable of
probing fundamental questions of ultra-high energy neutrino
interactions. Disagreement with the Standard Model prediction for
the cross section could be an indication of new physics. The problem
is that the knowledge of neutrino flux and neutrino-nucleon cross
section must be built up simultaneously, since we are largely
ignorant of both in the energy regime of interest. The knowledge of
one is dependent on knowledge of the other. In these conditions we
have defined in a previous work \cite{alfa} an observable
($\alpha(E)$) to search effects of new physic in the
neutrino-nucleon interaction. This new observable works by comparing
the surviving flux at the detector such that the observable is
weakly dependent of the initial flux.

\begin{figure*}
\includegraphics[angle=270,width=2.7in,bb= 180 180 580 680]{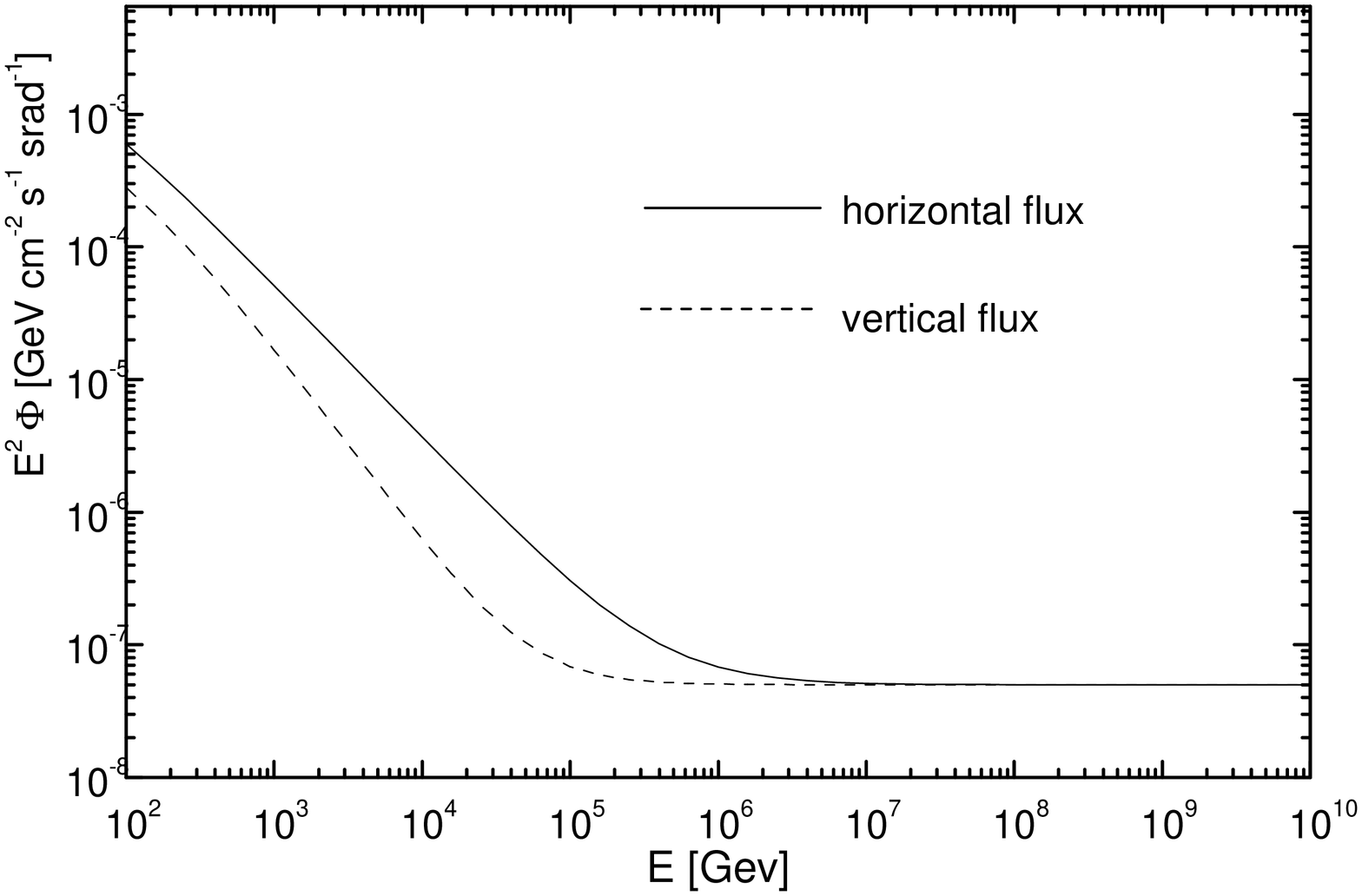}
\caption{\label{fig:flux0}  The utilized flux obtained by adding the
atmospheric and the isotropical cosmic flux. }
\end{figure*}

The angle $\alpha(E)$ and the related ratio $\eta(E)$ introduced in
Ref.\cite{alfa} are the observable that we shall use in this paper
in order  to study the impact of leptoquark physics on neutrino
detection in a neutrino telescope such as IceCube. By definition
$\alpha(E)$ is the angle that divides the Earth into two homo-event
sectors. When neutrinos traverse the planet in their journey to the
detector, they find different matter densities, and then, different
number of nucleons to interact with. In this conditions, the number
of neutrinos that finally arrive to the detector depends on the
arrival directions, indicated by the angle $\theta$ with respect to
the nadir direction. If we consider only upward-going neutrinos of a
given energy $E$, that is, the ones with arrival directions $\theta$
such that $0<\theta<\pi/2$, there will always exist an angle
 $\alpha(E)$ such that the number of events for
$0<\theta<\alpha(E)$ equals that for $\alpha(E)<\theta<\pi/2$.

Clearly, the value of $\alpha(E)$ is energy dependent. For low
energies, the cross section decreases and the Earth becomes
transparent to neutrinos. In this case $\alpha(E)\rightarrow \pi/3$
for a diffuse isotropic flux since this angle divides the hemisphere
into two sectors with the same solid angle. Obviously for extremely
high energies, where most neutrinos are absorbed,
$\alpha(E)\rightarrow \pi/2$, and for intermediate energies
$\alpha(E)$ varies accordingly between these limiting behaviors.

In order to define $\alpha(E)$ we consider the expected number of
events (muon tracks though charged currents $\nu_{\mu}N$
interactions) at IceCube in the energy interval $\Delta E$ and in
the angular interval $\Delta \theta$ that can be estimated as
\begin{equation}\label{numberevent}
{\mathcal N}=n_{\rm T} T \int_{\Delta\theta}\int_{\Delta E} d\Omega
dE_{\nu} \sigma^{CC}(E) \Phi(E,\theta),
\end{equation}
where $n_{\rm T}$ is the number of target nucleons in the effective
detection volume, $T$ is the running time, and $\sigma^{CC}(E)$ is
the charged neutrino-nucleon cross section. We take  the detection
volume for the events equal to the instrumented volume for IceCube,
which is roughly 1 km$^3$ and corresponds to $n_{\rm T}\simeq 6
\times 10^{38}$. The function $\Phi(E,\theta)$  in
Eq.(\ref{numberevent}) is the survival flux which is the solution
(Eq.(\ref{transporte})) of the complete transport equation
\cite{nicolaidis}.

\begin{figure}[!t]
\centering
\includegraphics[angle=270,width=3.in,bb= 180 180 550 680]{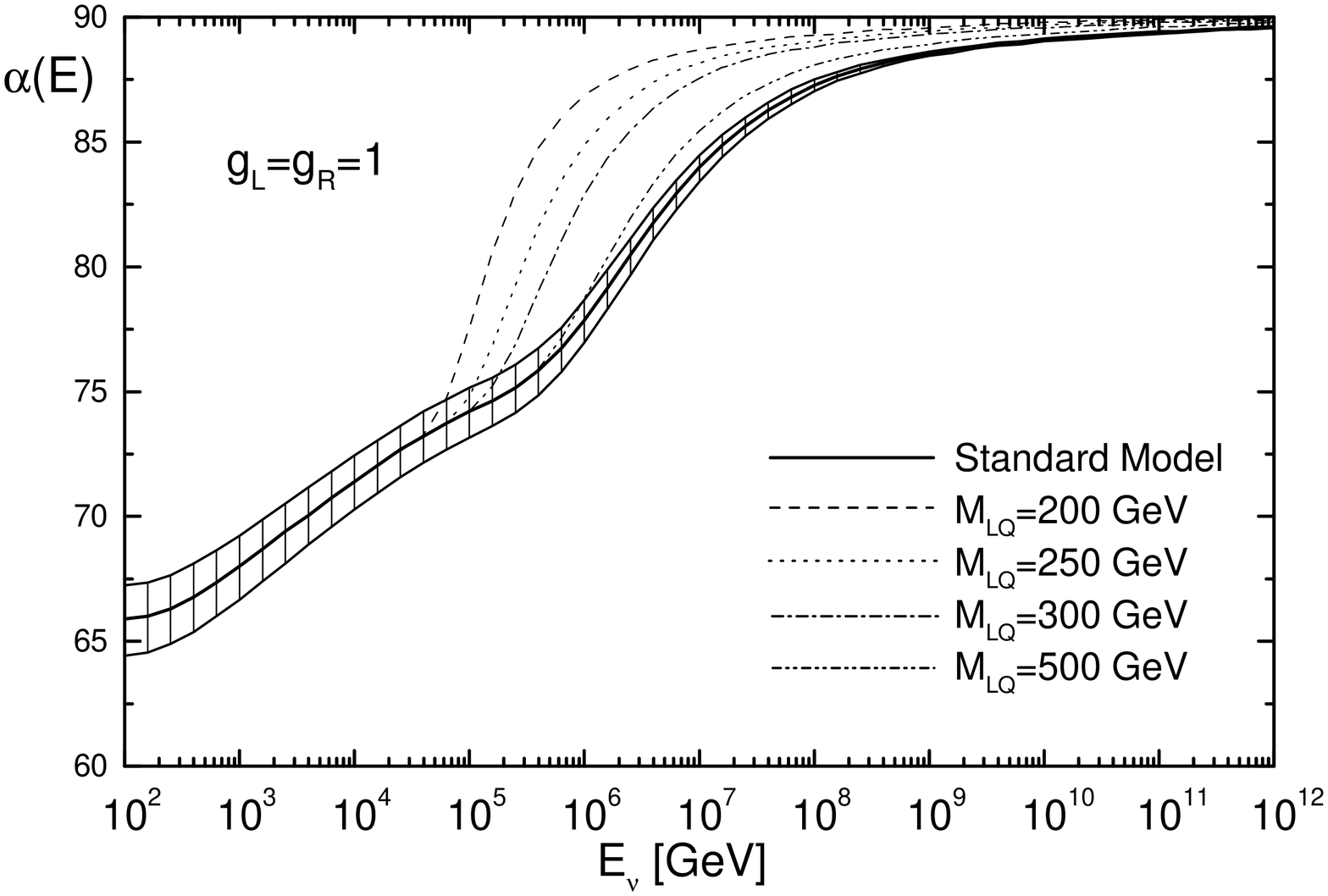}
\caption{\label{fig:alfainthelq}  The predictions for $\alpha(E)$
obtained for different values of $M_{LQ}$. The shaded region
represent the theoretical uncertainties for $\alpha_{SM}(E)$.}
\end{figure}

The definition of $\alpha(E)$ is essentially the equality between
two number of events, thus, to a good approximation, for each energy
bin all the previous factors cancel except the integrated fluxes at
each side. In this way, $\alpha(E)$ can be defined by the equation
\begin{equation}\label{alfadef}
\int_0^{\alpha(E)}d\theta \sin\theta \; \Phi_0(E,\theta)
e^{-\sigma_{eff}(E,\theta) \tau(\theta)}=\int_{\alpha(E)}^{\pi/2}
d\theta \sin\theta \; \Phi_0(E,\theta) e^{-\sigma_{eff}(E,\theta)
\tau(\theta)},
\end{equation}
which is numerically solved to give the results shown in the
Fig.~\ref{fig:alfainthelq}(solid line). There we show the SM
prediction for $\alpha(E)$, the theoretical uncertainties as we
explain below and the leptoquarks contribution for different values
of the mass $M_{LQ}$ and for the coupling $g=g_L=g_R=1$.

The main characteristics of $\alpha(E)$ have been reported recently
in Ref.\cite{alfa}. It is worth to notice that $\alpha(E)$ is weakly
dependent on the initial flux but, at the same time it is strongly
dependent on the neutrino nucleon cross-section. Hence, the use of
the observable $\alpha(E)$ reduces the effects of the experimental
systematics and initial flux dependence. Since the functional form
of $\alpha(E)$ sharply depends on the interaction cross section
neutrino-nucleon, if physics beyond the SM operates at these high
energies it will become manifest directly onto the function
$\alpha(E)$.

In order to evaluate the impact of the observable $\alpha(E)$ to
bound new physics effects, we have estimated the corresponding
uncertainties on the SM prediction for $\alpha(E)$. Considering the
number of events as distributed according to a Poisson distribution
the uncertainty can be propagated onto the angle $\alpha_{\rm
SM}(E)$. The number of events $N$ as a function of $\alpha_{\rm SM}$
is
\begin{equation}
N=2 \pi n_{\rm T} T \Delta E \sigma^{CC}(E) \int_0^{\alpha_{\rm
SM}}d\theta \sin\theta \; \Phi_0(E,\theta)
e^{-\sigma_{eff}(E,\theta)\tau(\theta)},
\end{equation}
where we have considered the effective volume for contained events
so that an accurate and simultaneous determination of the muon
energy and shower energy is possible and then of the neutrino
energy. For IceCube, it corresponds to the instrumented volume,
roughly 1 km$^3$, implying a number of target nucleons $n_{\rm
T}\simeq 6 \times 10^{38}$. We have considered an integration time
$T=15$ yr which is the expected lifetime of the experiment.
%If we propagate the error for $N$, the error with which $\alpha$ is
%obtained into the one in
To propagate the error on $N$ to obtain the one on $\alpha$, we note
that
\begin{equation}
\delta N=\dfrac{dN}{d\alpha} \delta\alpha,
\end{equation}
and dividing by $N$ we obtain the {\it statistical errors} on
$\alpha$
\begin{equation} \label{deltaalfa}
\delta\alpha=\left[\int^{\alpha_{\rm SM}(E)}_0 d\theta \left(
\dfrac{\sin\theta}{\sin\alpha_{\rm SM}}
\right)\left(\dfrac{\Phi_0(E,\theta)}{\Phi_0(E,\alpha_{SM})} \right)
\dfrac{\exp{[-\sigma_{eff}(E,\theta)\tau(\theta)]}}{\exp{[-\sigma_{eff}(E,\alpha_{SM})\tau(\alpha_{SM})]}}\right]
\left( \dfrac{\delta N}{N} \right),
\end{equation}
where for Poisson distributed events we have
\begin{equation}
\delta N=\sqrt{N}.
\end{equation}

\begin{figure}[!t]
\includegraphics[angle=270,width=3.in,bb= 180
180 550 680]{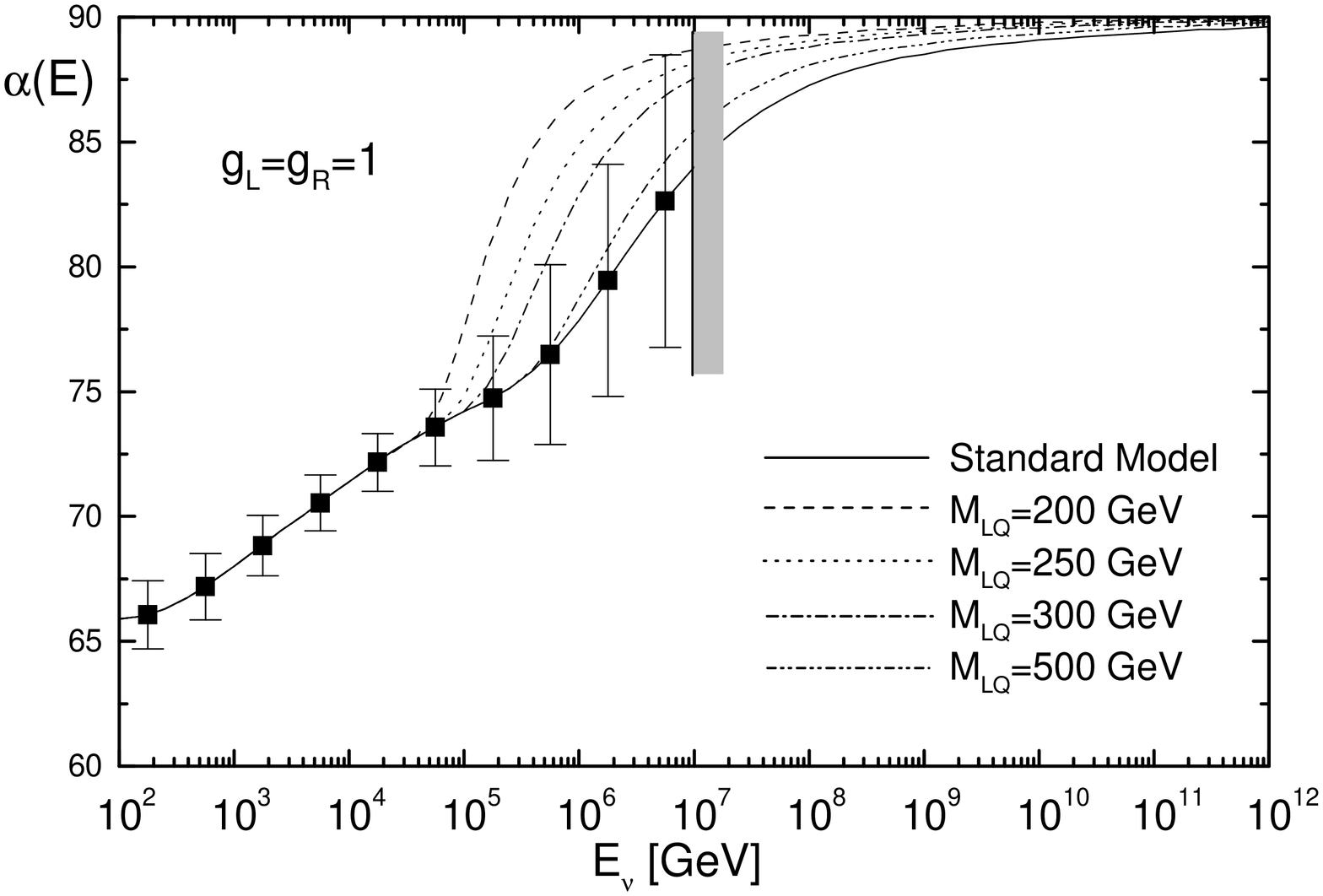}
\caption{\label{fig:alfaerrtotallq}  The predictions for $\alpha(E)$
obtained for different values of $M_{LQ}$. The indicated errors are
obtained by adding in quadrature the theoretical uncertainties and
the statistical errors as it just was explained in the text.}
\end{figure}

In order to evaluate the {\it errors} on $\alpha(E)$, it is
necessary to consider a level of initial flux $\Phi_0(E,\theta)$.
Here we have added together the cosmological diffuse isotropic flux
and the atmospheric one(see Fig.~\ref{fig:flux0}).  For the
atmospheric flux, we have considered the one given in
Ref.\cite{atmos}. For the cosmological diffuse flux, the usual
benchmark is the so-called Waxman-Bahcall (WB) flux for each flavor,
$E_{\nu_{\mu}}^2 \phi_{\rm WB}^{\nu_{\mu}}\simeq 2.4\times 10^{-8}
{\rm GeV} \ {\rm cm}^{-2} {\rm s}^{-1} {\rm sr} ^{-1}$, which is
derived assuming that neutrinos come from transparent cosmic ray
sources \cite{waxman-bahcall}, and that there is an adequate
transfer of energy to pions following $pp$ collisions. However, one
should keep in mind that if there are in fact hidden sources which
are opaque to ultra-high energy cosmic rays, then the expected
neutrino flux will be higher.

On the other hand, we have the experimental bounds set by AMANDA. A
summary of these bounds can be found in
Refs.\cite{desiati,amanda-bound} and as a representative value we
take $E_{\nu_{\mu}}^2 \phi_{\rm AM}^{\nu_{\mu}}\simeq  8 \times
10^{-8} {\rm GeV} \ {\rm cm}^{-2} {\rm s}^{-1} {\rm sr} ^{-1}$. With
the intention to estimate the number of events, we have considered
an intermediate flux (INT) level slightly below the present
experimental bound by AMANDA,
\begin{equation}
E_{\nu_{\mu}}^2 \phi_{\rm INT}^{\nu_{\mu}}\simeq 5 \times 10^{-8}
{\rm GeV} \ {\rm cm}^{-2} {\rm s}^{-1} {\rm sr} ^{-1}.
\end{equation}

Moreover we have considered the theoretical uncertainties on the
observable $\alpha_{SM}(E)$, which come from the uncertainties in
the earth density, the standard model neutrino-nucleon cross section
and the initial neutrino flux. As we explained above, we have
considered the initial flux as the sum between the atmospheric flux
and an isotropic diffuse cosmic flux. The observable $\alpha(E)$ is
weakly dependent on the isotropic uncertainties in the cosmic flux,
but it is dependent on the anisotropic uncertainties in the
atmospheric flux. Uncertainties in the calculated neutrino intensity
arise from lack precise knowledge of the input quantities, which are
the primary spectrum and the inclusive cross section for production
of pions and kaons by hadronic interaction in the atmosphere. If we
consider the primary spectrum as isotropic then, the corresponding
uncertainties do not affect significatively to $\alpha(E)$. Any
isotropic overall factor that we include to modify the initial
atmospheric flux do not produce higher effects on $\alpha(E)$
because it is defined by comparing the number of events from
different angular directions. In this conditions the principal
source of uncertainties is the inclusive cross section for
production of pions and kaons. We include theoretical uncertainties
in the energy-angular dependence in the atmospheric flux due to the
uncertainties in the $K/\pi$ ratio as an angular uncertainty between
horizontal and vertical neutrino events. In order to take into
account these uncertainties and their effect on $\alpha(E)$ we have
multiplied the atmospheric initial flux by an angular dependent
factor, imposing opposite uncertainties of $\pm$10\% to horizontal
and vertical flux respectively and interpolating for intermediate
angular values. In a similar way we have taken into account the
uncertainties that come from the earth density and the
neutrino-nucleon cross section \cite{premm,saakar}, such that it
maximize the uncertainties on $\alpha(E)$. The results are shown as
shaded region in Fig.\ref{fig:alfainthelq}.

We consider the theoretical uncertainties discussed above and the
statistical errors (Eq.(\ref{deltaalfa})) as uncorrelated
statistical errors and we sum them in quadrature. The results are
shown as error bars in Fig.(\ref{fig:alfaerrtotallq}).

As it was discussed in Ref.\cite{alfa} the interval for maximum
sensitivity for $\alpha$ is $10^5 \rm{GeV} <E<10^7 \rm{GeV}$.
However, as for lower energies the atmospheric flux grows and then
the statistical errors fall, we have considered as an energy window
for the fits the interval: $10^2 \rm{GeV} <E<10^7 \rm{GeV}$. In
Fig.~\ref{fig:alfaerrtotallq} we show our results for the observable
$\alpha(E)$ and the corresponding errors within the mentioned energy
window. In Fig.~\ref{fig:flux0} we show the used flux.

%%%%%%%%%%%%%%%%%%%%%%%%%%%%%%%%%%%%%%%%%%%%%%%%%%%%%%%%%%%%%%%%%%

\begin{figure*}
\centering
\includegraphics[scale=1.,angle=270,width=3.in,bb= 180 180 580 680]{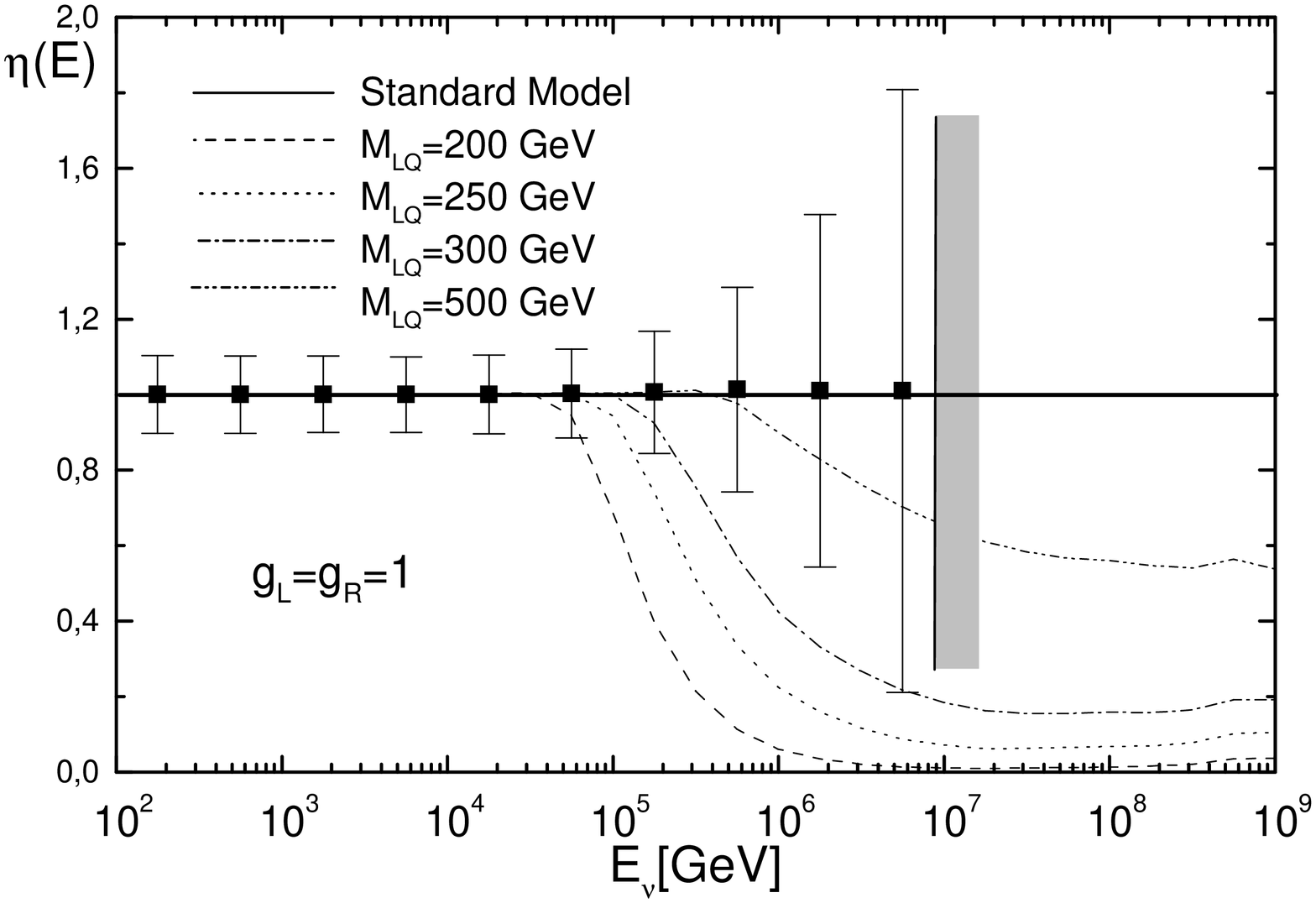}
\caption{\label{fig:etaerrtotallq} $\eta(E)$ for different values of
$M_{LQ}$. We include the statistical errors obtained of a number of
events distributed as a Poisson distribution added in quadrature
with the theoretical uncertainties as it was explained in the text.}
\end{figure*}

 In the same context, we can define another
observable related to $\alpha(E)$. We consider the hemisphere
$0<\theta<\pi/2$ divided into two regions by the angle
$\alpha_{SM}(E)$, ${\cal R}_1$ for $0<\theta<\alpha_{SM}(E)$ and
${\cal R}_2$ for $\alpha_{SM}(E)<\theta<\pi/2$. We then calculate
the ratio $\eta(E)$ between the number of events for each region,
\begin{equation}\label{eta}
\eta(E)=\frac{N_1}{N_2},
\end{equation}
where $N_1$ is the number of events in the region ${\cal R}_1$ and
$N_2$ the number of events in the region ${\cal R}_2$. By using
$\eta(E)$ the effects of experimental systematic and initial flux
dependence are reduced. If there is only SM physics, then we have
that the ratio $\eta_{SM}(E)=1$. In order to estimate the capability
of $\eta(E)$ to bound leptoquarks effects, we have considered the
values of $\eta(E)$ along with their error bars in
Fig.~\ref{fig:etaerrtotallq} as if they had been obtained from
experimental measurements for $\eta(E)$. We proceed, then, to
perform a $\chi^2$-analysis taking as free parameters the leptoquark
mass $M_{LQ}$ and the couplings $g=g_L=g_R$ and considering as {\it
experimental point} the SM values for $\eta(E)$ for the same energy
bin used in Fig.~\ref{fig:alfaerrtotallq}. We define the $\chi^2$
function in the usual way,
\begin{equation}
\chi^2=\sum_{i=1,10} \dfrac{(\eta_{\rm
SM}(E_i)-\eta(E_i,M_{LQ},g))^2}{(\delta\eta(E_i))^2}.
\end{equation}

\begin{figure*}
\centering
\includegraphics[scale=1.,angle=270,width=3.in,bb= 180 180 580 680]{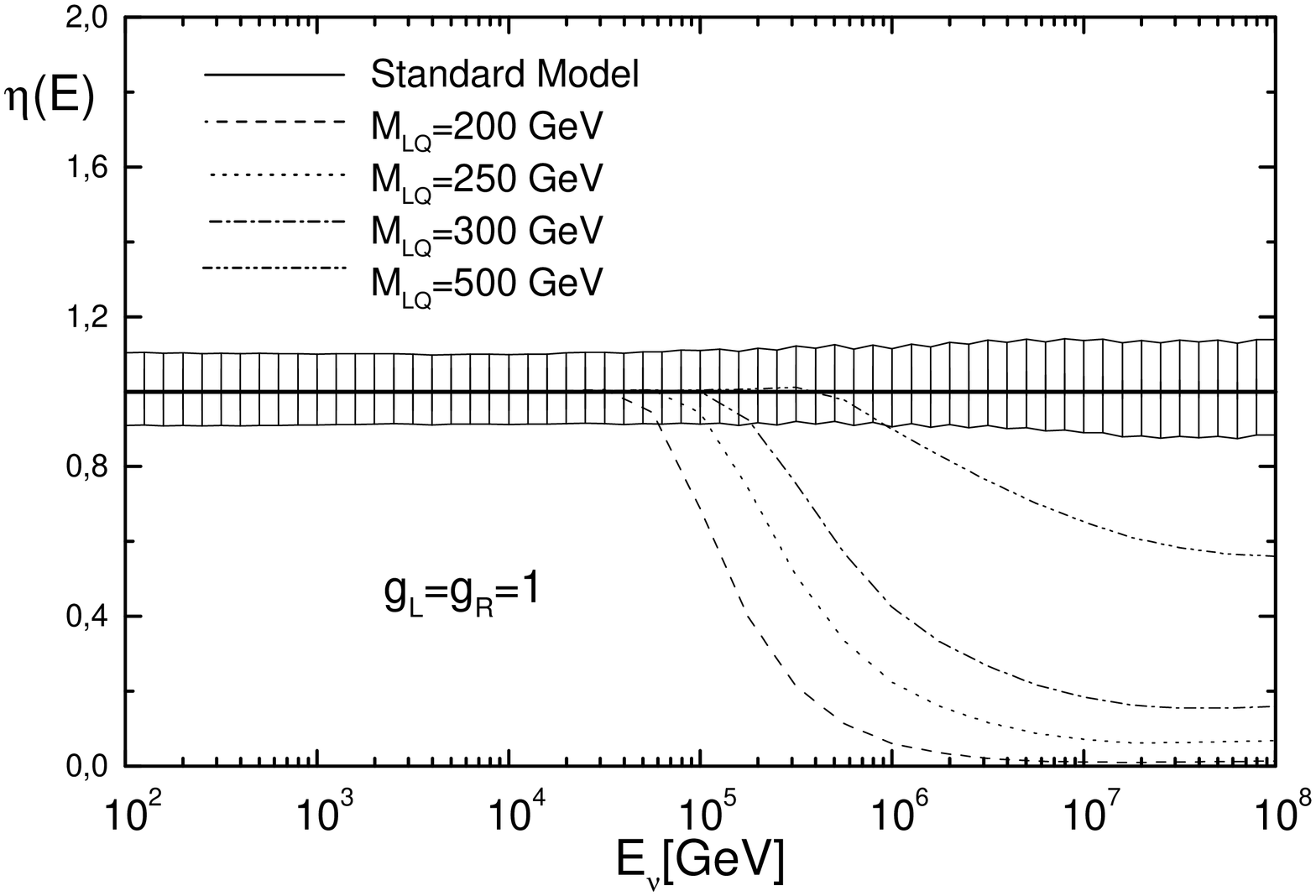}
\caption{\label{fig:etainthelq} $\eta(E)$ for different values of
$M_{LQ}$. We include the theoretical uncertainties as a shaded
region around the standard model value $\eta=1$.}
\end{figure*}

where $\delta\eta$ are errors obtained by adding in quadrature the
statistical errors and the theoretical uncertainties. According to
the definition of $\eta(E)$ (Eq.(\ref{eta})) the statistical errors
are given by $\delta\eta_{st}(E_i)=\sqrt{2/N_i}$ for events
distributed according to a Poisson distribution. In the same way
that we have done for $\alpha(E)$ we can propagate the theoretical
uncertainties on the observable $\eta(E)$ and these are show in
Fig.(\ref{fig:etainthelq}) as a shaded region around the Standard
Model prediction ($\eta=1$). These theoretical uncertainties are
added in quadrature with the statistical errors for $\eta(E)$ and
the results are show as errors bar in the
Fig(\ref{fig:etaerrtotallq}). Is important to realize that the
atmospheric flux is lower than the cosmic one for energies higher
than $10^5 GeV$. In the other hand the leptoquark contribution is
very small for energies lower than $10^5 GeV$. In this conditions we
do not expect a strong dependence of the leptoquarks bounds on the
atmospheric flux uncertainties.

\begin{figure*}
\begin{center}
\includegraphics[scale=1.,angle=270,width=3.in,bb= 100 180 580 680
]{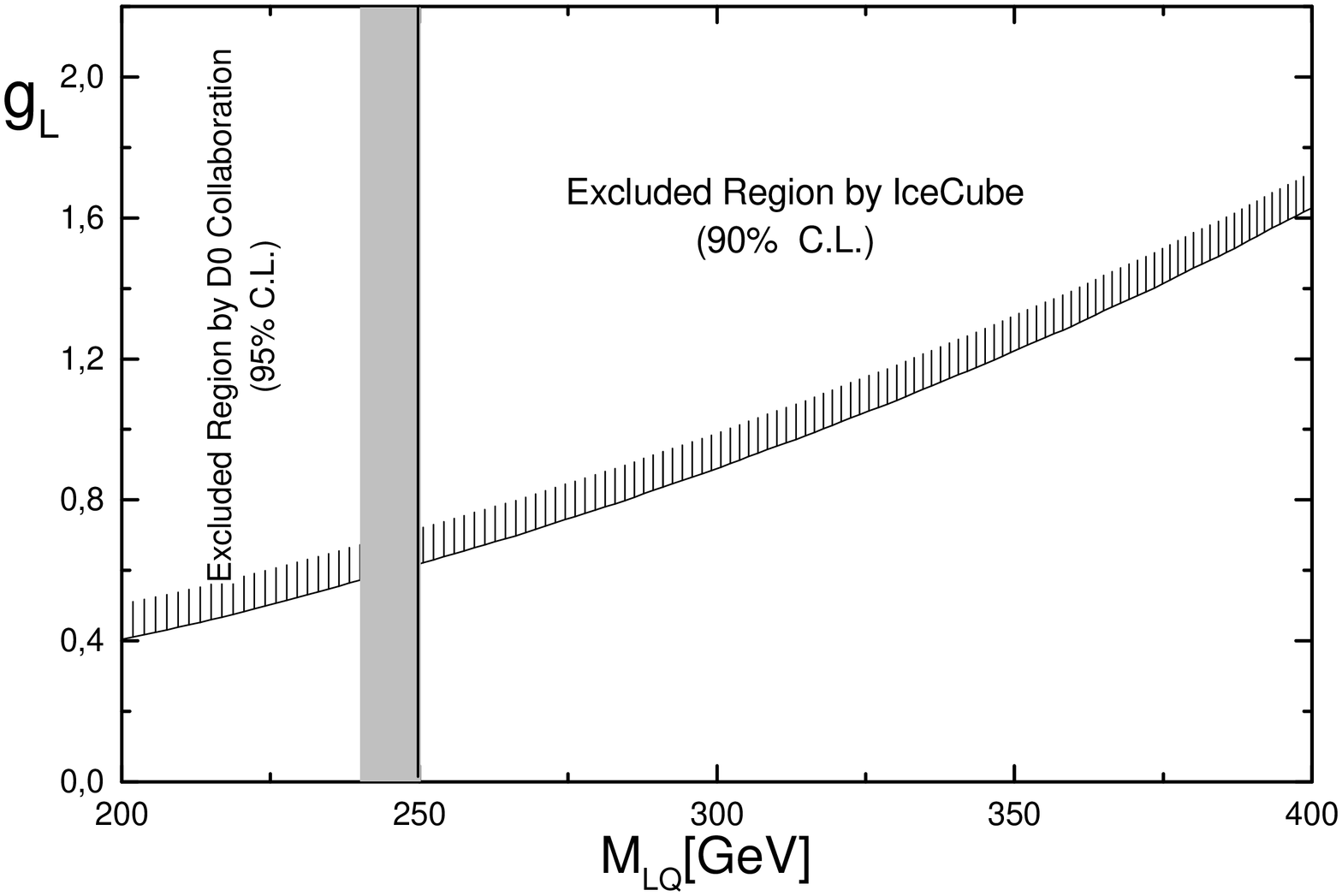} \caption{\label{fig:regerrtotallq} The
excluded regions in the $(M_{LQ},g)$ plane (for $g_L=g_R=g$) is
above the curve. The region to left the vertical is excluded by D0
\cite{d0}.}
\end{center}
\end{figure*}

The function $\chi^2$ is minimized to obtain the allowed region in
the ($M_{LQ}$, $g$) plane for $g_L=g_R=g$, which corresponds to the
region below the curve shown in Fig.~\ref{fig:regerrtotallq}. In the
same figure we also include the bounds obtained from the D0
experiment obtained for the second family (vertical line) \cite{d0}.

\section{Conclusions}
In the present work we studied the effects of leptoquarks
contributions to the neutrino-nucleon cross section on the survival
neutrino flux in a neutrino telescope like IceCube.  We have  found
a considerable disagreement with the SM prediction for the neutrino
observables defined above, particulary for low values of $M_{LQ}$.
For high values of $M_{LQ}$ this disagreement tends to disappear.

We have  also studied the possibility to bound effects of
leptoquarks contributions to the interactions between muon neutrinos
and the nucleons of the Earth using the observable $\eta(E)$.  In
this context, we fitted the theoretical expression for $\eta(E)$ as
a function of the $M_{LQ}$ and $g=g_L=g_R$ taking as experimental
data the SM values obtained for $\eta$ ($\eta_{SM}(E)=1$) along with
the errors that come from the theoretical uncertainties and the
number of events distributed according to a Poisson distribution.
The results are shown in Fig.~\ref{fig:regerrtotallq} as a allowed
region plot. Finally, would like comment that a similar region was
obtained in Ref. \cite{leptogarcia}, but using the down-going
neutrinos and the inelasticity distribution of events as an useful
observable also defined in IceCube. Perhaps, the simultaneous use of
both methods will make possible to improve the bounds on leptoquarks
physics.

\begin{acknowledgments}
We thank CONICET (Argentina) and Universidad Nacional de Mar del
Plata (Argentina).
\end{acknowledgments}

%\newpage

\end{document}